\documentclass[a4paper,12pt]{article}

\usepackage[english]{babel}
\usepackage[utf8x]{inputenc}
\usepackage[T1]{fontenc}
\usepackage{authblk}
\usepackage{natbib}
\usepackage{here}
\usepackage{setspace}
\usepackage{multirow}
\usepackage{lmodern}
\usepackage{amsmath, amssymb}
\usepackage{bm}
\usepackage{pdflscape}
\usepackage{authblk}
\usepackage{color}
\usepackage[table]{xcolor}

\usepackage{soul}

\usepackage{booktabs}
\usepackage{afterpage}
\usepackage{textcomp}
\usepackage{float}
\usepackage{enumerate}
\usepackage{rotating}

\definecolor{babyblue}{rgb}{0.54, 0.81, 0.94}
\definecolor{babypink}{rgb}{0.96, 0.76, 0.76}

\usepackage[margin=1in]{geometry}

\usepackage{amsmath}
\usepackage{graphicx}
\usepackage[colorlinks=true, allcolors=blue]{hyperref}

\usepackage{caption}
\usepackage{subcaption}

\usepackage{gensymb}

\begin{document}

    \begin{center}
        \vspace*{1cm}
        \large
	    \textbf{Multivariate spatial models for small area estimation of species-specific forest inventory parameters}\\
         \normalsize
           \vspace{5mm}
	    Jeffrey W. Doser\textsuperscript{1}, Malcolm S. Itter\textsuperscript{2}, Grant M. Domke\textsuperscript{3}, Andrew O. Finley\textsuperscript{4, 5}
         \vspace{5mm}
    \end{center}
    \small
        \textsuperscript{1}Department of Forestry and Environmental Resources, North Carolina State University, Raleigh, NC, USA \\
        \textsuperscript{2}Department of Environmental Conservation, University of Massachusetts, Amherst, MA, USA. \\
        \textsuperscript{3}Northern Research Station, United States Department of Agriculture Forest Service, St. Paul, MN, USA.\\
        \textsuperscript{4}Department of Forestry, Michigan State University, East Lansing, MI, USA \\
         \textsuperscript{5}Department of Statistics and Probability, Michigan State University, East Lansing, MI, USA \\
        \noindent \textbf{Corresponding Author}: Jeffrey W. Doser, email: jwdoser@ncsu.edu.\\
         \normalsize

\section*{Abstract}

National Forest Inventories (NFIs) provide statistically reliable information on forest resources at national and other large spatial scales. As forest management and conservation needs become increasingly complex, NFIs are being called upon to provide forest parameter estimates at spatial scales smaller than current design-based estimation procedures can provide. This is particularly true when estimates are desired by species or species groups, which is often required to inform wildlife habitat management, sustainable forestry certifications, or timber product assessments. Here we propose a multivariate spatial model for small area estimation of species-specific forest inventory parameters. The hierarchical Bayesian modeling framework accounts for key complexities in species-specific forest inventory data, such as zero-inflation, correlations among species, and residual spatial autocorrelation. Importantly, by fitting the model directly to the individual plot-level data, the framework enables estimates of species-level forest parameters, with associated uncertainty, across any user-defined small area of interest. A simulation study revealed minimal bias and higher accuracy of the proposed model-based approach compared to the design-based estimator. We applied the model to estimate species-specific county-level aboveground biomass for the 20 most abundant tree species in the southern United States using Forest Inventory and Analysis (FIA) data. County-level biomass estimates from the proposed model had high correlations with design-based estimates, yet the model-based estimates tended to have a slight positive bias relative to design-based estimates, particularly for abundant and managed species. Importantly, the proposed model provided large gains in precision across all 20 species. On average across species, 91.5\% of county-level biomass estimates had higher precision compared to the design-based estimates. Future work should explore incorporation of additional auxiliary data sources that can help explain fine-scale variation in biomass of managed species. The proposed framework is an attractive solution for NFI data users to generate species-level forest parameter estimates with reasonable precision at management-relevant spatial scales.  

\textbf{Keywords}: Forest Inventory and Analysis; National Forest Inventory; Bayesian; model-based inference.

\newpage

\section{Introduction}

Monitoring forest resources across broad spatial and temporal scales is increasingly important for forest management to meet both economic and sustainability objectives. National forest inventories (NFIs) are designed to estimate forest parameters on regional to national scales using a large network of permanent inventory plots selected according to a probability sampling design \citep{mcroberts2010advances}. For example, the United States (US) Department of Agriculture Forest Service Forest Inventory and Analysis (FIA) program conducts the US NFI, which generates a valuable resource for quantifying forest health, timber resources, and forest landowner characteristics across broad regions of the US \citep{bechtold2005enhanced}. NFIs, like the FIA program, historically use design-based inference to estimate forest parameters, which generally provides sufficiently precise forest parameter estimates across large areas of forestland (e.g., US states) \citep{westfall2022sampling}. However, NFI programs across the world have experienced increased demand for estimates within smaller spatial, temporal, and biophysical extents than design-based inference and existing sampling intensities can reasonably deliver \citep{breidenbach2012small, prisley2021needs}. Methodological approaches that provide sufficiently precise estimates within small areas of interest using data from a larger sample are collectively referred to as small area estimation (SAE; \citealt{rao2015small}). Such techniques have seen increasing applications with NFI data in recent years (e.g., \citealt{hou2021updating, cao2022increased, finley2024models, shannon2025toward}).

Both design-based and model-based inferential paradigms have been employed in SAE applications with NFI data \citep{dettmann2022review}. Regardless of the inferential paradigm used, most SAE approaches seek to relate the variable of interest to auxiliary information from satellite and airborne remote sensing data through a statistical model. When a strong relationship exists between the auxiliary information and variable of interest, the precision of small area estimates can be substantially improved \citep{sarndal2003model}. In addition to incorporating auxiliary information through regression-based frameworks, model-based SAE approaches can explicitly borrow information from outside the area of interest through the use of random effects. Random effect structures can range from simple, normally distributed effects to account for similarities in forest parameters across ecologically-similar areas (e.g., ecoregions), management units, or political boundaries (e.g., \citealt{white2021hierarchical}), to more complex forms that explicitly acknowledge spatial and spatial-temporal dependencies in forest parameters through a spatial covariance structure (e.g., \citealt{finley2024models,shannon2025toward}). Recent advances in model-based SAE approaches that accommodate spatial dependence have been shown to provide improved precision of forest parameters and forest owner characteristics at state \citep{ver2017hierarchical, harris2021sae}, regional \citep{cao2022increased}, and national levels \citep{may2023spatially, shannon2025toward}.  

While considerable progress has been made in developing SAE methods suitable for forest inventory, a vast majority of proposed methods focus on estimating a single forest attribute such as total merchantable volume or aboveground biomass \citep{dettmann2022review}. Estimates of forest attributes by species or species class are often of interest due to individual species representing different forest products and values, providing unique benefits to wildlife, and serving as useful indicators of site quality and/or disturbance history \citep{kershaw2016forest}. While independent models can be fit for each individual species, such an approach fails to leverage dependencies between species-specific attributes, which can substantially improve precision of model predictions given the importance of species interactions (e.g., competition) in determining stand structure and dynamics \citep{weiskittel2011forest, clark2014more, tang2025zero}. Total biomass within a fixed area is limited by site quality, and as a result the biomass of one species generally cannot increase without a corresponding decrease in the biomass of another \citep{clark2017generalized}. Additionally, reliable small area estimates of species-specific forest inventory parameters are complicated by large amounts of zero-inflation and highly right skewed data distributions. For example, species tend to be absent from many plots, present at low abundances in many plots, and present at high abundances in only a few plots. In many cases, individual species are never observed in the small number of plots within an area of interest, which prevents the calculation of a standard error estimate using design-based estimation. Furthermore, even for small areas with some non-zero values of species-specific attributes, the distribution of values is typically highly right-skewed and close to zero, which calls into question the reliability of confidence intervals that rely on the central limit theorem. Thus, there is a need for model-based frameworks that deliver precise estimates of forest parameters by species and similar high-dimensional forest attributes. 

Here we develop a multivariate spatial model for SAE of species-specific forest inventory parameters. Our hierarchical Bayesian model simultaneously accounts for zero-inflation in individual species responses, correlations among species, and residual spatial autocorrelation---factors that notoriously complicate analysis of forest inventory data (e.g., \citealt{finley2012hierarchical, clark2014more, may2023spatially}). We implement the model at the unit-level (i.e., FIA plot-level), which provides flexibility in generating species-specific small area estimates across user-defined areal units of interest. We perform a simulation study to assess bias and accuracy of the model relative to alternative design-based  estimators. The proposed model is applied to FIA data in the southern US to estimate county-level aboveground live and standing dead tree biomass for the 20 most abundant tree species in the region. The multivariate model generates county-level estimates with substantially improved precision compared to alternative design-based estimators, showcasing the utility of our approach to generate species-level forest parameter estimates from NFI data at management-relevant spatial scales. 

\section{Materials and methods}

\subsection{FIA NFI data}

FIA operates a nationally-consistent, quasi-systematic survey designed to monitor forest resources across the US. Around 1999, the FIA program transitioned to a systematic grid of permanent sample plots that are evenly distributed in an annual panel system to provide statistically robust estimates of both current forest resources and change over time. Each individual panel consists of a subset of the permanent sample plots measured in the same year that represent complete spatial coverage across the population of interest. Inventory cycles in the southern US consist of 5 or 7 annual panels. The base intensity of permanent ground plots across the continental US is approximately 1 plot per 2,428 hectares \citep{smith2002forest, bechtold2005enhanced}.

On forested FIA plots, trees with a diameter at breast height (d.b.h) $\geq 12.7$ cm are measured on a cluster of four circular 0.02 ha subplots centered on each plot location \citep{bechtold2005enhanced}. Within each subplot, trees 2.54-12.70 cm d.b.h. are measured on a circular 13.5 m$^2$ microplot. Tree attributes (e.g., species, alive/dead/harvested status) and quantitative variables (e.g., d.b.h., height, volume) are measured following nationally-standardized protocols. Aboveground biomass for each measured tree is estimated following FIA's National Scale Volume and Biomass Estimators (NSVB) framework \citep{westfall2024national, johnson2025}. These individual tree estimates are summarized and expanded to provide a plot-level estimate of aboveground biomass (Mg/ha). 

\begin{figure}[!h]
    \centering
    \includegraphics[width=12cm]{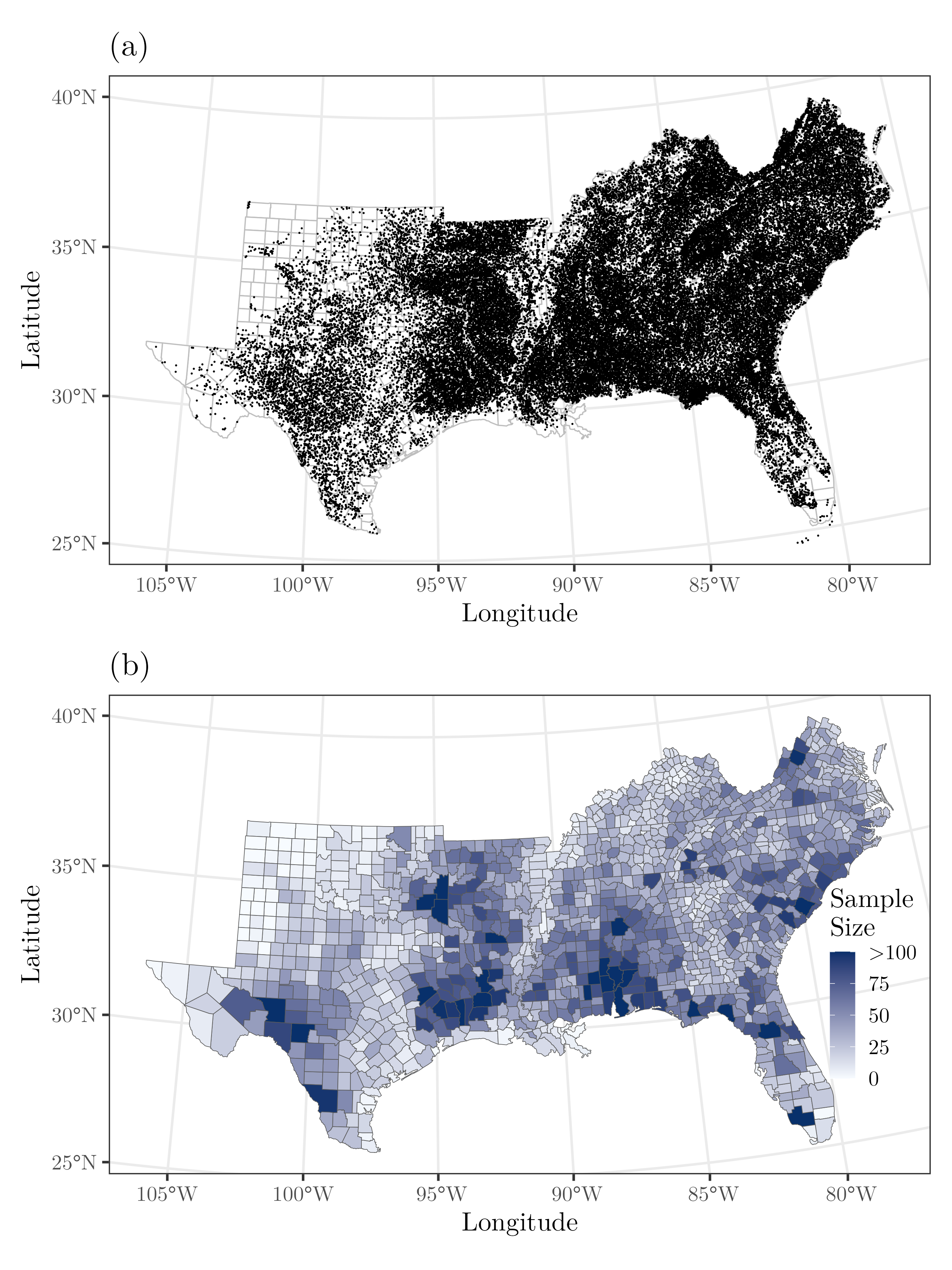}
    \caption{Plot locations (a) and distribution of $n = 46,710$ Forest Inventory and Analysis (FIA) plots across the 1306 counties (small areas; b) in the southern United States. Note the plot locations are the publicly available perturbed locations in which FIA adds a small amount of random noise to the true plot locations.}
    \label{fig:fiaPlots}
\end{figure}

Here we use plot-level aboveground biomass data from $n = 46,710$ FIA plots across the FIA Southern Region of the United States (Figure \ref{fig:fiaPlots}). We only use the most recent measurement at a given inventory plot and only use those measurements taken since 2015. We use the publicly available perturbed plot locations in which FIA adds a small amount of random noise to the true plot locations to protect private plot locations and preserve ecological integrity of the plots. Our focus is on estimating species-specific aboveground biomass at the county-level for all live and dead stems (i.e., snags leaning less than 45 degrees) $\geq 2.54$cm d.b.h. of the twenty most abundant species in the region (Table \ref{tab:modelCor}). We used the \texttt{rFIA} package \citep{stanke2020rfia, rFIA} in \texttt{R} \citep{R} to download and extract the plot-level FIA data.

\subsection{FIA sampling design and design-based estimators}

The annual FIA sampling design implemented in the late 1990s is based on a systematic tessellation of $\sim$2400 ha hexagons superimposed over the US based on a random starting location \citep{bechtold2005enhanced, westfall2022sampling}. Plot locations within each hexagon were determined following a pseudo-random approach. For cells with an existing inventory plot from previous iterations of the FIA program, a set of predefined rules were used to select a single plot. For cells without any pre-existing plots, plot locations were selected randomly within each cell. Grid cells were then assigned to annual panels to ensure the subset of panels sampled in any given year provided approximately uniform spatial coverage of the population. 

As is commonly done in forest inventory, FIA uses post-stratified design-based estimators to estimate forest parameters of interest in an attempt to reduce variance of direct estimates without post-stratification \citep{westfall2011post}. Post-stratification proceeds by using wall-to-wall digital imagery to classify locations into a set of strata (e.g., forest/nonforest, canopy cover proportion classes). Stratum weights are then calculated based on the proportion of all pixels in the population assigned to the given stratum, and population estimates are obtained as a weighted average of the stratum-specific estimates (see \citealt{westfall2022sampling} for a full description of FIA post-stratification techniques). While useful for estimating national and regional level forest parameters, such post-stratified estimators are limited in SAE applications, particularly when generating estimates by species due to individual stratum not being present within each SAE, highly zero-inflated and right-skewed distributions of species-specific forest parameters, and very small sample sizes within each stratum/SAE combination. Accordingly, for comparison with our model-based estimates, we use a simple design-based direct estimator (i.e., the sample mean). More specifically, letting $y_{i, j, k}$ denote the aboveground biomass for species $j$ in plot $i$ that falls in county $k$, our design-based estimate for county-level biomass is 

\begin{equation}\label{directMu}
  \hat{\mu}_{j, k} = \frac{1}{n_k}\sum_{i = 1}^{n_k}y_{i, j, k},
\end{equation}

where $n_k$ is the number of inventory plots in county $k$. The associated variance for (\ref{directMu}) is 

\begin{equation}\label{directVar}
  \hat{\sigma}^2_{j, k} = \frac{1}{n_{k}(n_{k} - 1)}\sum_{i = 1}^{n_k}(y_{i, j, k} - \hat{\mu}_{j, k})^2. 
\end{equation}

\subsection{Auxiliary data}

We extracted climate, forest cover, and remotely-sensed elevation data for use as auxiliary information in the subsequently described SAE framework. We extracted all climate variables from TerraClimate \citep{abatzoglou2018terraclimate}, which provides high-resolution ($1/24$ degree or approximately $4$km) climate normals from 1981-2010 across the globe. In particular, we extracted monthly climate normals for maximum temperature, minimum temperature, total precipitation, and vapor pressure deficit. From these monthly values, we calculated maximum temperature (TMAX) as the maximum of the monthly maximum temperature normals, minimum temperature (TMIN) as the minimum of the monthly minimum temperature normals, total precipitation (PPT) as the sum of the monthly precipitation normals, and vapor pressure deficit (VPD) as the average of the growing season (April through October) monthly vapor pressure deficit normals. We calculated tree canopy cover (TCC) at each FIA plot center in 2019 using the National Land Cover Database (NLCD) 2021 Forest Service Science data set \citep{houseman2023national}. Finally, we determined the elevation at the center of each FIA plot using bare-earth terrain height data from Amazon Web Service Terrain Tiles (accessed on 13 June 2025 from \url{https://registry.opendata.aws/terrain-tiles}).

\subsection{Modeling framework}\label{model}

We seek a modeling framework that can provide improved precision of species-specific forest parameter estimates by incorporating plot-level predictor variables correlated with aboveground biomass, exploiting correlations between co-occurring species, and borrowing information from spatially proximate plots. We accomplish this via a multivariate log-normal hurdle model that accommodates the large zero-inflation and heavily right-skewed distribution of species-level biomass. The proposed model first jointly estimates the presence or absence of each species across the set of $n = 46,710$ FIA plots to establish the distribution of each species across the region. We subsequently estimate species-specific biomass only at those plots where the individual species is present. We refer to these sub-models as Stage 1 and Stage 2, respectively. Estimates from the model are then used to predict species-specific biomass across a dense grid of cells over the region of interest, which we ultimately summarize to generate species-level estimates across the 1,306 counties in the southern US. In the subsequent model presentation, we focus on the case of estimating species-specific biomass in the southern US, but note that the model could be used to predict any species-specific forest parameter that takes positive-only values (e.g., basal area, volume, carbon). We used the following R packages for data downloading, processing, and manipulating auxiliary data: \texttt{sf} \citep{pebesma2023spatial}, \texttt{tidyverse} \citep{wickham2019welcome}, \texttt{climateR} \citep{climateR}, and \texttt{elevatr} \citep{elevatr}.

\subsubsection*{Stage 1: Binary sub-model}

We first jointly estimate the presence or absence of all species to accommodate the large propensity of locations where a species does not occur (and thus has zero biomass). Let $\bm{s}_i$ denote the spatial coordinates of the $i^{\text{th}}$ forest plot for each of the $i = 1, \dots, 46,710$ plots (e.g., the easting and northing coordinate for the center subplot of a FIA inventory plot $i$). Define $z_j(\bm{s}_i)$ as the presence (1) or absence (0) of species $j$ at plot $i$ for each of the $j = 1, \dots, J$ species, where here $J = 20$. We model $z_j(\bm{s}_i)$ following
\begin{equation}\label{yBern}
  z_j(\bm{s}_i) \sim \text{Bernoulli}(\psi_j(\bm{s}_i))\;,
\end{equation}
where $\psi_{j}(\bm{s}_i)$ is the probability species $j$ occurs at plot $i$. We model $\psi_{j}(\bm{s}_i)$ as a function of multiple climate variables and a spatially varying intercept. More specifically, we have 
\begin{equation}\label{psiEq}
  \begin{split}
  \text{logit}(\psi_j(\bm{s}_i)) &= \beta_{0, j} + \text{w}^\ast_{j}(\bm{s}_i) + \beta_{1, j} \cdot \text{TMIN}_i + \beta_{2, j} \cdot \text{TMIN}^2_i + \\
  &\beta_{3, j} \cdot \text{TMAX}_i + \beta_{4, j} \cdot \text{TMAX}^2_i + \beta_{5, j} \cdot \text{PPT}_i + \beta_{6, j} \cdot \text{PPT}^2_i, 
  \end{split}
\end{equation}
where $\beta_{0, j}$ is the overall intercept for species $j$, $\beta_{1, j}$ and $\beta_{2, j}$ are linear and quadratic effects of minimum temperature (TMIN), $\beta_{3, j}$ and $\beta_{4, j}$ are linear and quadratic effects of maximum temperature (TMAX), $\beta_{5, j}$ and $\beta_{6, j}$ are linear and quadratic effects of total precipitation, and $\text{w}^\ast_{j}(\bm{s}_i)$ is a spatially varying intercept for species $j$ at plot $i$ to accommodate residual spatial structure in the species distribution. We included minimum/maximum temperature and total precipitation to help establish individual species distributions given their importance in determining species ranges across the southern US. We include linear and quadratic effects to allow the relationships to peak at intermediate levels of one or more of the climate variables in relation to the climatic niche of the individual species. 

We model all regression coefficients and the non-spatial intercept for each species hierarchically (i.e., as random effects; \citealt{dorazio2005, gelfand2005modelling}). More specifically, for each $t = 0, \dots, p$, we model $\beta_{j, t}$ according to 
\begin{equation}\label{betaEq}
  \beta_{j, t} \sim \text{Normal}(\mu_{\beta_t}, \tau^2_{\beta_t})\;,    
\end{equation}
where $\mu_{\beta_t}$ is the mean effect across all species, and $\tau^2_{\beta_t}$ is the variability in the effect across all species. By treating species-specific effects as random effects, we borrow strength across species and obtain more precise estimates for both rare and common species \citep{zipkin2009impacts}.

We seek to jointly model the species-specific spatially varying intercepts to account for residual correlations between species and their occurrence patterns. We use a spatial factor modeling approach \citep{hogan2004bayesian}, a dimension reduction technique that can efficiently model correlations among a large number of species. This approach has seen recent applications in forest inventory for integrating highly-multivariate LiDAR data with classical forest inventory data \citep{taylor2019spatial}, ultimately providing substantial gains in computational run time than alternative multivariate modeling approaches (e.g., the linear model of coregionalization). We decompose $\text{w}^*_{j}(\bm{s}_i)$ into a linear combination of $q$ latent variables (i.e., factors) and their associated species-specific coefficients (i.e., factor loadings). More specifically, we have
\begin{equation}\label{wStarEq}
  \text{w}^*_{j}(\bm{s}_i) = \bm{\lambda}_{j}^\top\textbf{w}(\bm{s}_{i})\;,
\end{equation} 
where $\bm{\lambda}_{j}^\top$ is a coefficient vector for species $j$ for each of the $q$ independent spatial factors at plot $i$ ($\textbf{w}(\bm{s}_{i})$). When $q$ is small relative to the total number of species, this approach provides substantial computational improvements compared to estimating a separate spatial process for each species \citep{tikhonov2020computationally, doser2023joint}. Such an approach accounts for species correlations not accounted for by the covariates using individual responses (i.e., loadings) to the $q$ spatial factors such that co-occurring species will have similar species-specific factor loadings \citep{doser2023joint}. 

The standard spatial factor approach predicts the spatial factors using a spatial Gaussian process (GP), which requires calculating the inverse and determinant of a dense $n \times n$ matrix. This computation involves $O(n^3)$ computations (floating point operations or FLOPs) for each of the $q$ spatial factors, which quickly renders the approach impractical for even modest sized forest inventory data sets (e.g., hundreds of locations). Here we estimate the latent spatial factors using a GP approximation called the Nearest Neighbor Gaussian Process (NNGP; \citealt{datta2016hierarchical, finley2019nngp}) that provides substantial improvements in run time, with negligible differences in inference and prediction, compared to a model that uses a full GP. For each $r = 1, \dots, q$ spatial factor, we have 
\begin{equation}\label{wEq}
    \text{w}_{r}(\bm{s}_i) \sim N(\bm{0}, \tilde{\bm{C}}_{r}(\phi_{r}))\;,
\end{equation}
 
where $\tilde{\bm{C}}_{r}(\phi_{r})$ is the NNGP-derived correlation matrix for the $r^{\text{th}}$ spatial process. We use an exponential spatial correlation model, where $\phi_{r}$ is a spatial decay parameter controlling the range of the spatial dependence for the $r$th spatial factor. The number of factors can be chosen using an exploratory approach where models with differing numbers of factors are used, compared using some model assessment criteria (e.g., Widely Applicable Information Criterion), and the number of factors can be set to the value where the model improvement becomes negligible (analogous to a scree plot in principal components analysis). For this analysis, we found $q = 5$ provided a sufficient balance between model fit and model run time. 

To complete the Bayesian specification of the model, we use normal priors with mean equal to 0 and variance set to 2.72 for all mean parameters, which results in a vague prior on the probability scale \citep{northrup2018comment}. We use vague inverse gamma priors with both the shape and scale parameter set to 0.1 for all variance parameters. We set moderately informative uniform priors to the spatial decay parameters such that the effective spatial range (i.e., the distance at which the correlation between two sites falls to 0.05) is bounded between 50km and 2000km. This allows for sufficient flexibility to estimate fine-scale and/or broad-scale spatial autocorrelation. Finally, we assign standard normal priors to the lower triangle of the factor loadings matrix. Following \cite{taylor2019spatial}, we fix the diagonal elements of the factor loadings matrix to 1 and the upper triangle to 0 to ensure identifiability of the factor loadings and the spatial factors. 

\subsubsection*{Stage 2: Log-normal sub-model}

For the second component of the hurdle model, we jointly estimate the species-specific biomass only at the locations where the species is observed to be present. Let $y_j(\bm{s}_i)$ denote the aboveground biomass for species $j$ at plot $i$. Analogous to \cite{finley2012hierarchical}, we predict $y_j(\bm{s}_i)$ conditional on the presence ($z_j(\bm{s}_i) = 1$) or absence ($z_j(\bm{s}_i) = 0$) of species $j$ at plot $i$. More specifically, when the species is present at plot $i$, we have
\begin{equation}\label{yEq}
   \text{log}(y_j(\bm{s}_i)) \mid (z_j(\bm{s}_i) = 1) \sim \text{Normal}(\mu_j(\bm{s_i}), \tau^2_{j}),
\end{equation}
where $\mu_j(\bm{s}_i)$ is the average biomass (on the log scale) for species $j$ at plot $i$, and $\tau^2_{j}$ is the species-specific non-spatial residual error (e.g., measurement error, microscale variation). We log-transform the variable of interest given the large positive skew in the species-level biomass data (Supplemental Information S1: Figure S1), such that using a log-normal distribution allows us to better adhere to basic linear model assumptions and to ensure positive support upon back-transformation. We model $\mu_j(\bm{s}_i)$ according to 
\begin{equation}\label{muEq}
  \begin{split}
  \mu_j(\bm{s}_i) &= \alpha_{0, j} + \eta^\ast_j(\bm{s}_i) + \alpha_{1, j} \cdot \text{TCC}_i + \alpha_{2, j} \cdot \text{VPD}_i + \\
  &\alpha_{3, j} \cdot \text{PPT}_i + \alpha_{4, j} \cdot \text{ELEV}_i + \alpha_{5, j} \cdot \text{ELEV}^2_i, 
  \end{split}
\end{equation}
where $\alpha_{0, j}$ is the overall intercept for species $j$, $\alpha_{1, j}$ is the linear effect of percent tree canopy cover (TCC), $\alpha_{2, j}$ is the linear effect of average growing season vapor pressure deficit (VPD), $\alpha_{3, j}$ is the linear effect of total precipitation (PPT), and $\alpha_{4, j}$ and $\alpha_{5, j}$ are linear and quadratic effects of elevation (ELEV). We included vapor pressure deficit and total precipitation as covariates on expected biomass given their physiological importance for tree growth. Percent tree canopy cover was hypothesized to have a positive relationship with biomass, particularly with common species. Elevation was included given its correlation with distinct forest types across the southern US data. The species-specific spatially varying intercept ($\eta^\ast_j(\bm{s}_i)$) is modeled using the spatial factor NNGP approach described in (\ref{wStarEq}) and (\ref{wEq}). As in Stage 1, we fit the model with five spatial factors.

At sites where each species does not occur, (i.e., $z_{j}(\bm{s}_i) = 0$), we model the variable of interest following
\begin{equation}\label{y2Eq}
     y_j(\bm{s}_i) \mid (z_j(\bm{s}_i) = 0) \sim \text{Normal}(0, 0.0001).
\end{equation}
By definition, aboveground biomass of species $j$ is zero at sites where the species does not occur and thus the measurement is not taken. However, we cannot simply fix $y_j(\bm{s}_i)$ to zero, as this would result in a degenerate likelihood \citep{finley2012hierarchical} that prevents prediction across a grid to form subsequent small area estimates. This approach avoids such degeneracy, ultimately providing stable estimation properties while ensuring estimates of inventory parameters at sites where a species does not occur are essentially estimated at zero. 

As opposed to the hierarchical approach used in Stage 1, here we assign independent normal priors with a mean of 0 and variance of 10 to the species-specific regression coefficients, as preliminary analyses revealed the shared prior was not adequate to describe variation in the parameters across species. We assigned inverse gamma priors with the scale parameter set to 2 and shape parameter set to 1 to species-specific residual variance parameters. We use the same priors and identifiability constraints for the spatial decay parameters and factor loadings matrix as described for Stage 1.

\subsection{Model implementation}

We estimated all model parameters in a Bayesian framework using efficient Markov chain Monte Carlo (MCMC) algorithms. We implemented Stage 1 using the \texttt{sfJSDM} function in the \texttt{spOccupancy} R package \citep{doser2022spoccupancy}. We ran three chains of the Stage 1 model for 200,000 MCMC iterations with a burn-in period of 140,000 iterations and a thinning rate of 30, yielding 6,000 posterior samples. We implemented Stage 2 using the \texttt{sfMsAbund} function in the \texttt{spAbundance} R package \citep{doser2024spabundance}, running the model for three chains each of 100,000 MCMC iterations, a burn-in period of 40,000, and a thinning rate of 30, yielding 6,000 posterior samples. Convergence was assessed using visual assessment of traceplots and the potential scale reduction factor (i.e., $\hat{\text{R}}$, \citealt{brooks1998general}).   

\subsection{Prediction and small area estimation}

Our inferential objective is to estimate species-specific biomass density (Mg/ha) across the 1,306 counties in the southern US. We accomplish this task by predicting biomass for each species across a dense grid of $n_0$ prediction locations overlaid across the study region. For our application, we overlaid a $1 \times 1$km grid of points across the southern US, which resulted in a total of $n_0 = 2,198,845$ prediction locations with a median of $1,515$ prediction locations falling into a single county. Prediction at each location proceeds in two stages, analogous to the model fitting process, during which we first predict the presence or absence of each species at each prediction site, and subsequently predict the amount of biomass for each species that is predicted to be present at a site. Importantly, because there is uncertainty in the set of locations where a given species is predicted to be present, we must fully account for this uncertainty when predicting biomass to avoid inflating the precision of our resulting estimates \citep{finley2012hierarchical}. The Bayesian framework readily enables such uncertainty propagation via posterior predictive inference. 

Let $\bm{s}_0$ denote a generic location across the grid of $n_0$ prediction locations. For each MCMC sample $l$, we predict the presence (1) or absence (0) of species $j$ at site $\bm{s}_0$ following
\begin{equation}
  z^{(l)}_{j}(\bm{s}_0) \sim \text{Bernoulli}(\psi_{j}^{(l)}(\bm{s}_0)),
\end{equation}

where $\psi_j^{(l)}(\bm{s}_0)$ is the predicted occurrence probability at location $\bm{s}_0$ for species $j$ that is calculated according to (\ref{psiEq}) using the parameter estimates at MCMC iteration $l$ obtained from fitting the model to the observed data. The prediction of biomass for species $j$ at prediction location $\bm{s}_0$ and MCMC iteration $l$ is then estimated conditional on $z^{(l)}_{j}(\bm{s}_0)$ such that
\begin{equation}
  y^{(l)}_{j}(\bm{s}_0) \sim \left\{
    \begin{matrix}    
      \text{Normal}(0, 0.0001),\hfill &z^{(l)}_j(\bm{s}_0) = 0 \\
      \text{exp}(\text{Normal}(\mu_j^{(l)}(\bm{s}_0), \tau^{2, (l)}_j)),\hfill &z^{(l)}_j(\bm{s}_0) = 1, \end{matrix} \right.
\end{equation}

where $\mu_j^{(l)}(\bm{s}_0)$ is the predicted mean biomass (on the log scale) calculated according to (\ref{muEq}). Once biomass is predicted at each of the $m = 1, \dots, n_{0,k}$ prediction locations in county $k$, county-level biomass density (Mg/ha) is obtained by simply taking the average of all point-level predictions, i.e., 

\begin{equation} \label{saeMean}
  \bar{y}^{(l)}_{j, k} = \frac{1}{n_{0,k}}\sum_{m = 1}^{n_{0,k}}y^{(l)}_j(\bm{s}_m).
\end{equation}

This process results in a full posterior distribution of biomass density predictions for each species in each county. These samples can then be summarized using any measure of central tendency (e.g., median, mean) to yield the final model-based estimate, along with an associated measure of uncertainty (e.g., standard deviation, 95\% credible interval, coefficient of variation). 

\subsection{Model assessment and comparison to design-based estimates}

We compared county-level biomass estimates from the proposed multivariate spatial model to the design-based direct estimates (i.e., sample mean) previously defined in (\ref{directMu}). Ideally, the proposed model will provide small area estimates that show high correlation with the design-based direct estimates, but are more precise. To assess improvements in precision, we compared the design-based coefficient of variation (i.e., the standard error divided by the sample mean) to the model-based coefficient of variation (i.e., posterior standard deviation divided by the posterior median) for biomass density estimates for each species in each county. The design-based estimator and the Bayesian model-based approach rely on fundamentally different statistical paradigms, so this comparison reflects how each quantifies uncertainty for its estimate of the population parameter---in this case, species-specific biomass in each county. See \cite{gregoire1998design} for a more complete discussion regarding the comparison of model-based and design-based estimates. As an assessment of correspondence between the design-based and model-based estimates, we calculated simple correlation coefficients for each species. We additionally calculated the average bias between model-based county-level estimates and the design-based county-level estimates. Note that these bias estimates do not necessarily represent bias from the true, unknown species-specific county-level biomass values and rather reflect differences of the model-based estimate from the design-based estimates, which themselves could be biased.

We used four-fold cross-validation to provide an assessment of the proposed model's ability to predict species-level small area estimates in small areas that were not used to fit the model. We randomly split the 1,306 counties into four approximately equal sets. We subsequently fit the proposed model four times, where each model was fit using the inventory plots found within three of the four sets of counties. We subsequently predicted species-specific biomass density within the counties that were not used when fitting the model and estimated bias between the resulting model-based estimates and the design-based estimates. We averaged the estimated bias value for each species across the four hold-out sets for a single measure of bias for each individual species. Our approach to generating hold out sets based on counties aligned with our desire to produce county-level estimates, and thus provided a more realistic assessment of model predictive performance compared to randomly holding out individual inventory plots.   

\subsection{Simulation study}

In addition to model assessment in the FIA case study, we performed a simulation study to quantify the bias and accuracy of the proposed model-based approach to species-specific SAE. We simulated a single population of $41,501$ population units distributed across a $1.75\times1.75$ km grid over the state of North Carolina. Each population unit represents a potential sampling unit. We simulated biomass of ten species across all population units following the multivariate spatial model described in Section \ref{model}, where we first simulated the presence/absence of each species at each population unit conditional on the climate variables and four spatial factors, then subsequently simulated the biomass for each location where the species occurs conditional on climate, elevation, tree canopy cover, and four spatial factors. The spatial factors were simulated to represent a mix of fine-scale and broad-scale spatial autocorrelation, a frequent phenomenon in forest inventory data (e.g., \citealp{may2025spatial}). This simulation approach resulted in data with similar characteristics to the observed FIA data (e.g., highly zero-inflated and right-skewed), with species occurring on average at approximately $\sim29\%$ (minimum $2\%$ and maximum $87\%$) of the population units in the region.  

From this population, we sampled 100 replicate data sets each comprising $1000$ population units selected at random from the population. For each of the replicate data sets, we calculated species-specific average biomass (Mg/ha) in 100 small areas (i.e., counties) across the state using the proposed model and design-based estimator. For both estimation approaches, we calculated bias as the difference between the estimate and true population value for each species and county, as well as the root mean square error (RMSE) as a measure of average accuracy. Complete details on the simulation study are provided in Supplemental Information S2. 

\section{Results}

\begin{figure}[!h]
    \centering
    \includegraphics[width=15cm]{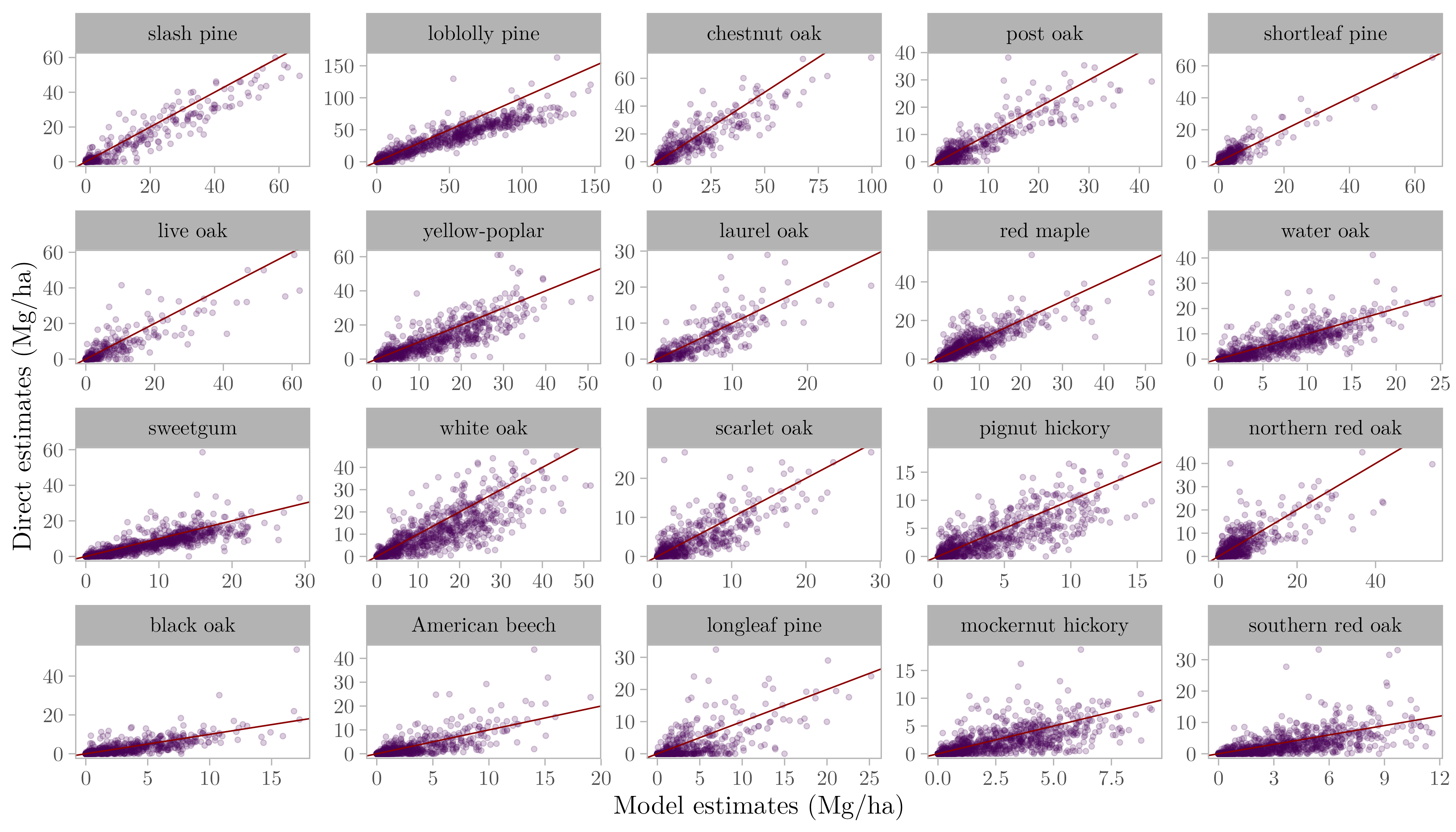}
    \caption{Comparison of small area estimates of species-specific biomass for the 20 species from the proposed multivariate spatial model and design-based direct estimates. Species are ordered in descending order of the correlation between direct and model-based estimates, with slash pine (\textit{Pinus elliottii}, top left) having the highest correlation and southern red oak (\textit{Quercus falcata var. falcata}, bottom right) having the lowest correlation.}
    \label{fig:overallComparisons}
\end{figure}

\begin{figure}[!h]
    \centering
    \includegraphics[width=15cm]{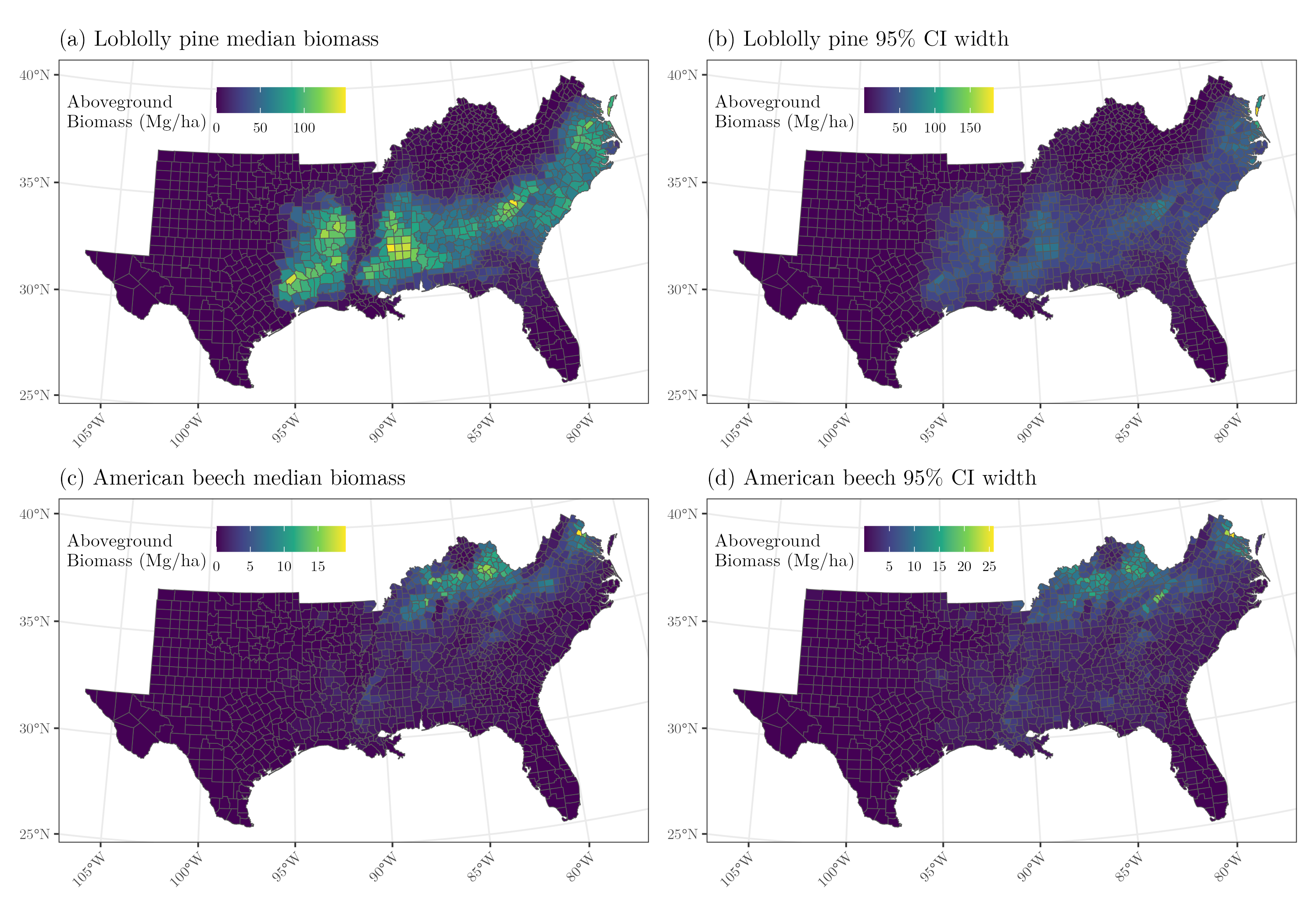}
    \caption{Model-based estimates (posterior medians) of aboveground biomass (Mg/ha; a, c) and associated 95\% credible intervals (b, d) for the  most abundant species (loblolly pine (\textit{Pinus taeda})), and least abundant species (American beech (\textit{Fagus grandifolia})) included in our analysis.}
    \label{fig:exampleMaps}
\end{figure}

Estimates of species-specific aboveground biomass from the multivariate spatial model across the 1,306 counties in the southern US showed close correspondence to design-based direct estimates across all twenty species (Figure \ref{fig:overallComparisons}). The average correlation coefficient between the model-based estimates and design-based estimates across the 20 species was 0.85, ranging from 0.96 for slash pine to 0.69 for southern red oak (Table \ref{tab:modelCor}). County-level estimates for the most abundant (loblolly pine) and least abundant (American beech) species are shown in Figure \ref{fig:exampleMaps}, along with associated 95\% credible intervals to characterize uncertainty in the estimates. Maps of estimates and their uncertainty for all 20 species are provided in Supplemental Information S1.  

\begin{figure}[!h]
    \centering
    \includegraphics[width=15cm]{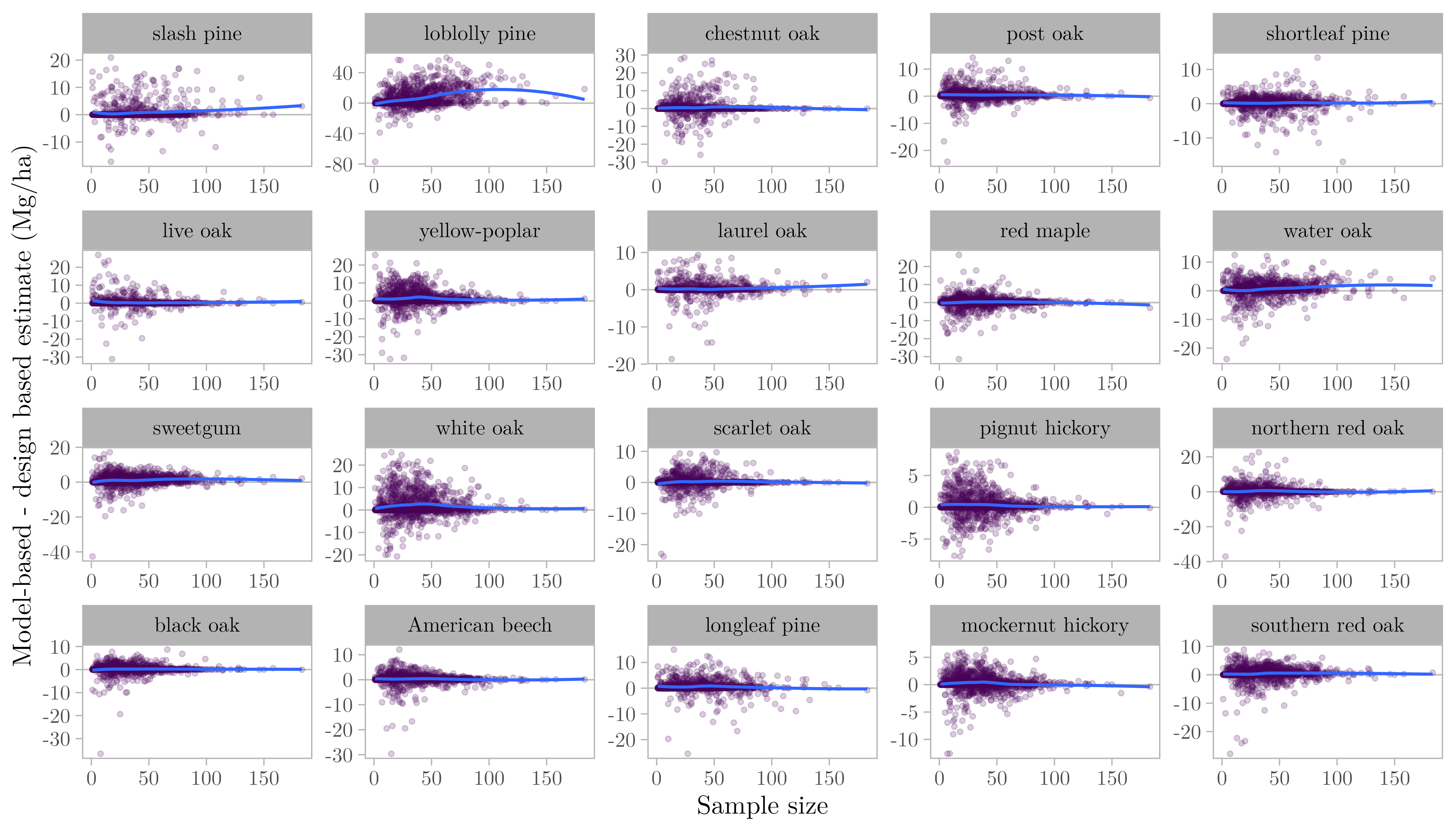}
    \caption{Relationship between sample size and bias in the model-based county-level estimates relative to the design-based estimates. Species are ordered in descending order of the correlation between direct and model-based estimates, with slash pine (top left) having the highest correlation and southern red oak (bottom right) having the lowest correlation. Note the sample sizes indicate all FIA plots in a county, regardless of whether the species was detected at the plot.}
    \label{fig:sampleSizeBias}
\end{figure}

\begin{figure}[!h]
    \centering
    \includegraphics[width=15cm]{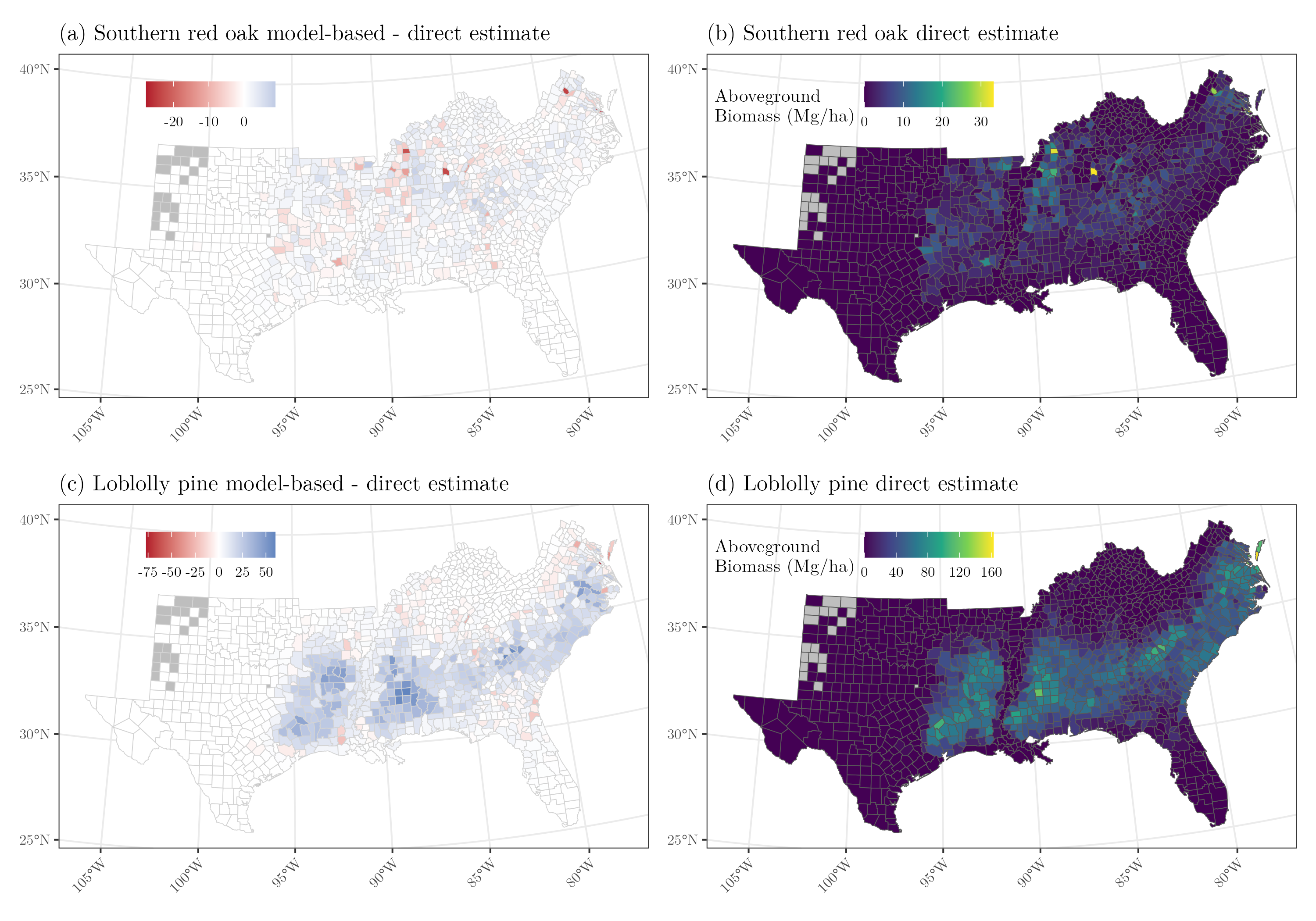}
    \caption{Differences between the design-based and model-based estimates for southern red oak (a) and loblolly pine (c) along with maps of the design-based direct estimate (b, d). Gray corresponds to counties with no FIA plots.}
    \label{fig:directCompare}
\end{figure}

Despite the correspondence between model-based and design-based estimates, model-based estimates had a slight positive bias in county-level biomass estimates relative to the design-based estimates (Table \ref{tab:modelCor}). Patterns in bias were often species-specific, with multiple species showing increases in bias as sample size increased (e.g., water oak, laurel oak, slash pine; Figure \ref{fig:sampleSizeBias}). Residual assessments revealed moderate deviations from normality in the form of a heavy left tail (Supplemental Information S3) for many species, with the deviations being the most substantial for species with the largest amount of positive bias in county-level estimates. In general, the model-based estimates tended to provide more spatially smooth estimates relative to fine-scale heterogeneity represented in the direct estimates, as would be expected given the sharing of information between plots across space via the spatial autocorrelation structure and auxiliary variables in the model-based estimates (Supplemental Information S4). Large differences between model-based and design-based estimates occurred as a result of: (1) outlier counties within the species primary range that had a very high direct estimate in which the model provided substantially lower estimates more in line with surrounding counties (e.g., Figure \ref{fig:directCompare}a,b); (2) locations with high biomass where the model-based estimates tended to extend estimates of high biomass farther in space than the direct estimates (e.g., Figure \ref{fig:directCompare}c,d); (3) spatially disparate counties with positive biomass according to the direct estimate but often substantially lower biomass predicted from the model-based estimate. These patterns of correspondence between model-based and design-based estimates with slight positive bias in the model-based estimates were consistent when considering the entire data set as well as the hold-out assessments from the four-fold cross-validation (Table \ref{tab:cvTable}). 

The model-based county-level estimates were consistently more precise across the twenty species compared to the design-based estimates. Across all species, 91.5\% of county-level biomass estimates had higher precision (in terms of the coefficient of variation) compared to the design-based estimates, with mockernut hickory (98.9\%) having the largest percentage improvements and longleaf pine (68.9\%) having the lowest percentage improvements (Table \ref{tab:modelCor}). Note that precision comparisons are only possible in counties where a species was observed on at least one FIA plot, as the design-based estimates do not have standard errors for counties where a species was not observed (i.e., all observations are zero).

\begin{figure}[!h]
    \centering
    \includegraphics[width=15cm]{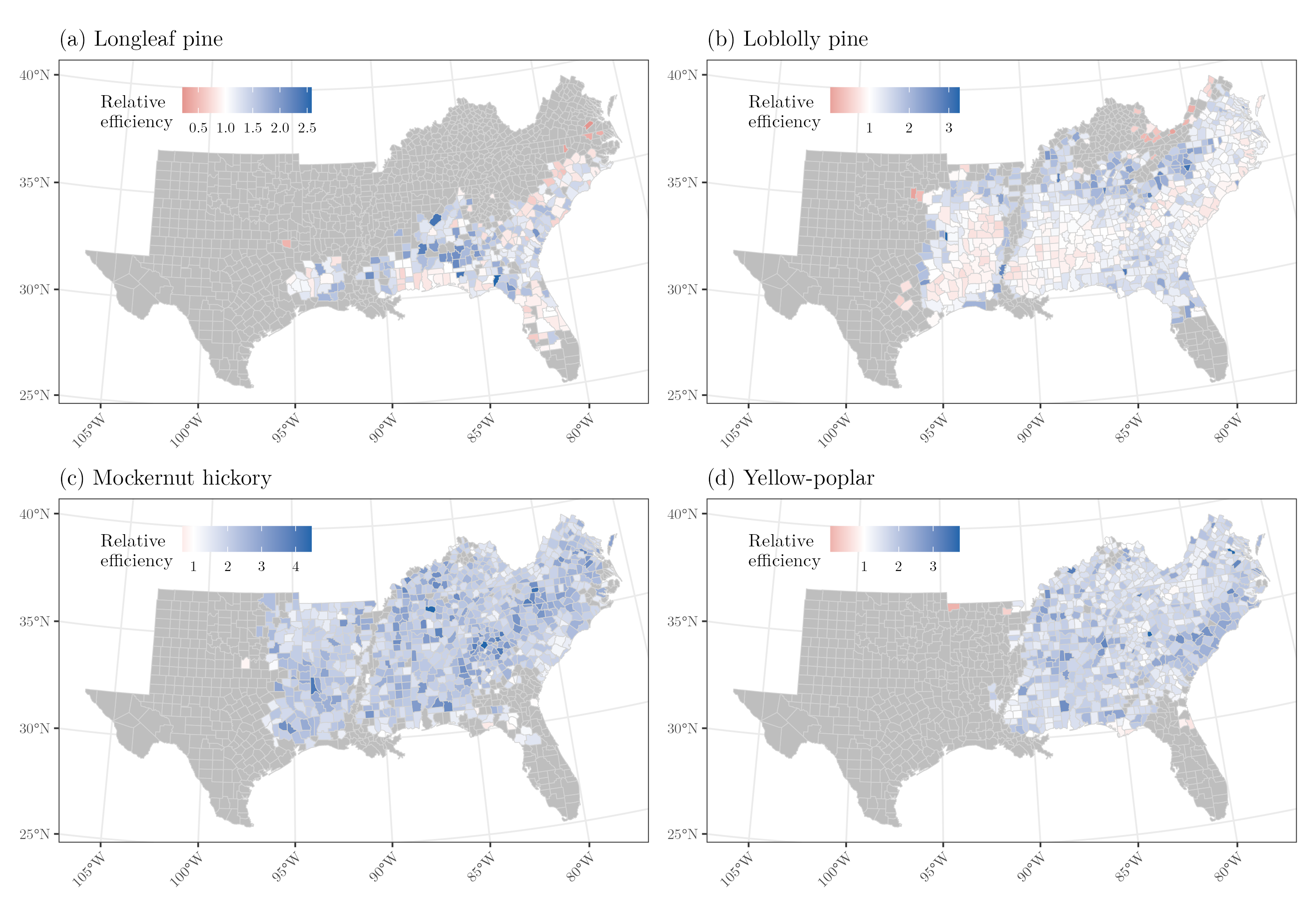}
    \caption{Relative efficiency of the model-based estimate coefficient of variation relative to the direct design-based estimate coefficient of variation for the two species with the smallest precision gains from the model-based approach (longleaf pine [\textit{Pinus palustris}] and loblolly pine [\textit{Pinus taeda}]) and the two species with the largest precision gains from the model-based approach (mockernut hickory [\textit{Carya tomentosa}] and water oak [\textit{Quercus nigra}]). Relative efficiency values greater than 1 (blue) indicate counties where the model-based estimate had higher precision, while values less than one (red) indicate counties where the model-based estimate had lower precision. Gray corresponds to counties where the design-based estimate is 0 and there is no design-based standard error.}
    \label{fig:cvMaps}
\end{figure}

\begin{figure}[!h]
    \centering
    \includegraphics[width=12cm]{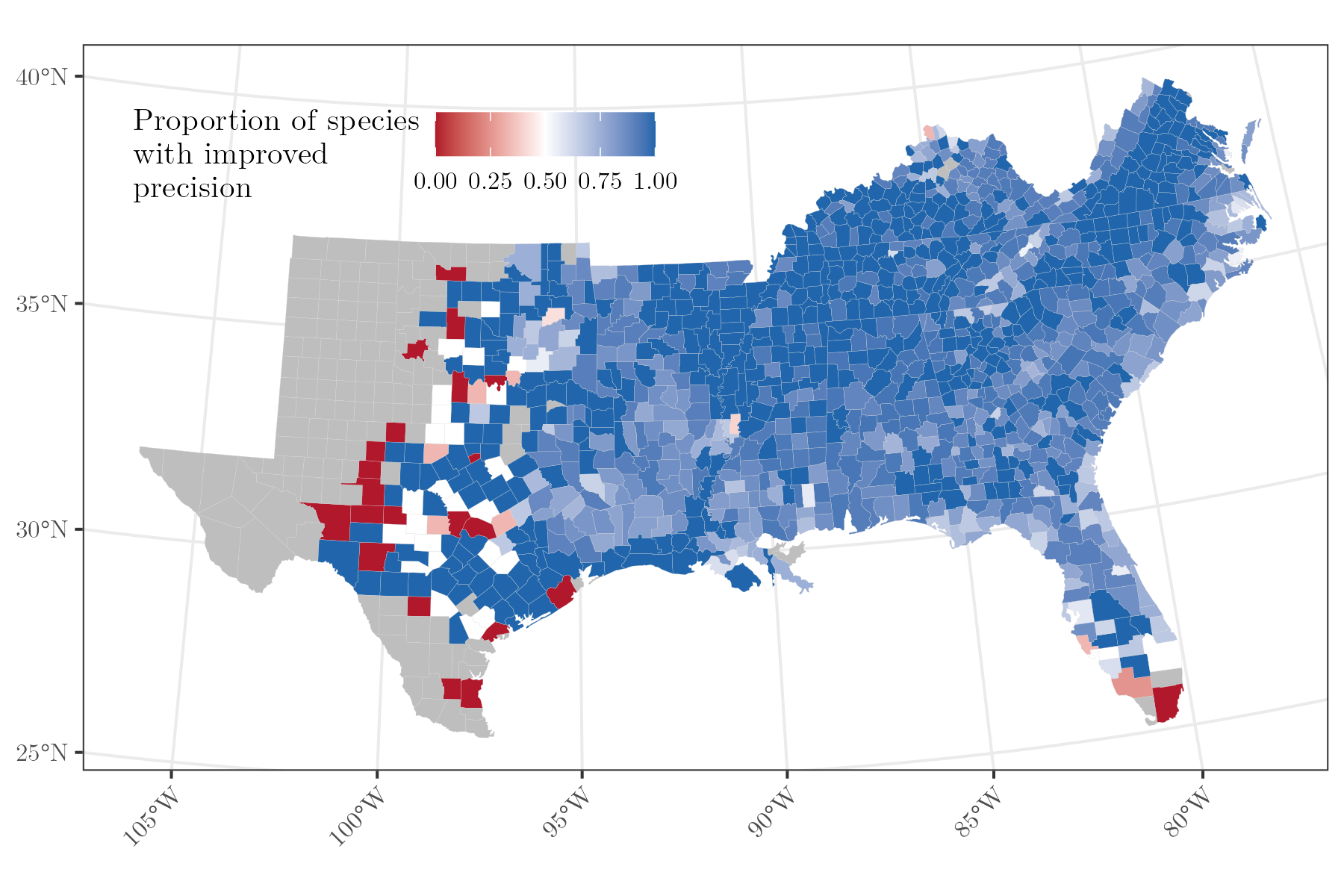}
    \caption{Proportion of species observed at least once in a given county for which the multivariate spatial model has higher precision relative to the design-based estimate. Gray corresponds to counties where none of the twenty species were observed.}
    \label{fig:allCVMap}
\end{figure}

Visualizing the relative efficiency of model-based county-level biomass estimates compared to the design-based direct estimates highlights the spatial extent of improvement the proposed model provides (Figures \ref{fig:cvMaps}, \ref{fig:allCVMap}). The multivariate spatial model led to widespread improvements across species in essentially the entirety of the southern US study region (Figure \ref{fig:allCVMap}). Noticeably, the multivariate spatial model had inferior performance to the design-based estimates in the western portion of the region as well as in southern Florida. Similar patterns emanate in maps of relative efficiency for individual species (Supplemental Information S1). For example, live oak, which has a complex spatial distribution spanning much of the southern and eastern coast of the study region, has large precision gains across Florida and much of Georgia, while precision is slightly decreased in much of the western portion of its range, particularly in several isolated counties (Supplemental Information S1: Figure S26). 

\subsection{Simulation study}

The proposed model-based estimator and the direct estimator did not have any clear directional bias across the ten species, with the sign of the bias varying across the ten species (Supplemental Information S2: Table 1). Variation in bias across counties tended to be larger for the model-based estimator than the direct estimator (Supplemental Information S2: Figure S1). Importantly, the model-based estimator had consistently higher accuracy (lower RMSE) compared to the direct estimator (Supplemental Information S2: Figure S2). 

\section{Discussion}

While NFIs have long provided statistically reliable information on forest resources at national and other large spatial scales, there is increased demand for estimates within smaller spatial scales \citep{dettmann2022review, prisley2021needs}. Standard design-based approaches are limited in their ability to provide reasonable precision of small area estimates, particularly when estimating forest parameters by species or species groups. Here we developed a multivariate spatial modeling framework to enable small area estimation of species-specific forest inventory parameters. In a simulation study, we found less bias and higher accuracy from our model compared to a design-based estimator. We applied our model to estimate county-level aboveground biomass for the 20 most abundant tree species in the southern US, which resulted in substantial precision gains relative to the design-based direct estimates. More broadly, our framework can be applied to estimate species-level forest parameters (e.g., biomass, carbon, volume) within management-relevant spatial scales to support diverse forest management objectives.  

County-level estimates of species-specific aboveground biomass in the southern US showed high correspondence with design-based estimates (Figure \ref{fig:overallComparisons}) and substantial improvements in estimate precision across the 20 species and 1,306 counties (Table \ref{tab:modelCor}, Figures \ref{fig:cvMaps}, \ref{fig:allCVMap}). Unlike the design-based direct estimator, the model-based approach can readily provide estimates for small areas without any inventory plots by leveraging information from auxiliary covariates and spatially proximate locations. Further, it is common for individual species to never be observed within the set of inventory plots in a given small area, particularly when the number of plot locations is small. While design-based estimates in such situations are always estimated at 0 with no corresponding standard error, the proposed model does not have such restrictions. The model-based estimator can readily provide uncertainty predictions in small areas where a species is never observed, and is also not forced to estimate the parameter at 0. This is particularly relevant for predicting attributes of rare species that might be missed in relatively small fixed area plots such as those used in FIA but occur in other portions of the small area of interest. In such a situation, the multivariate spatial model can use information from the auxiliary variables in the model, the spatial random effects, and the presence of other species to help estimate the forest parameter of interest.

Additionally, design-based confidence intervals require the estimator's sampling distribution to be approximately normally distributed. This is often argued to be achieved through the central limit theorem when sample sizes are sufficiently large. While some results show a sample size of $n = 25-30$ is sufficient for an approximately normal population \citep{hogg2020probability}, the distributions of species-specific forest inventory parameters are often highly skewed, zero-inflated, and non-symmetric (\citealt{tang2025zero}; Supplemental Information S1: Figure S1). Such data require even larger sample sizes for the normality assumption to be met, which is often not achievable for small area estimation, and thus confidence intervals for design-based small area estimates of species-specific parameters are likely not reliable.   

The application of the proposed model to species-specific biomass estimation in the southern US provides an important step forward for generating reliable species-specific small area estimates from FIA data. Individual species represent distinct values to wildlife, forest products, and functions for overall ecosystem health. Thus, reliable estimation of species-specific forest parameters from publicly available FIA data provides an important asset for forest managers and landowners for strategic planning, assessments of timber supply, and support for sustainable forestry certification. While we focused on estimation at the county-level, a key benefit of our proposed model is that it is implemented at the unit-level (i.e., inventory plot level), ultimately allowing users to generate small area estimates in any areal unit of interest larger than the individual plot size. Furthermore, the proposed model is not restricted to estimating average parameter values across a small area. Estimates of total biomass, carbon, or other variable of interest, with associated uncertainty, can readily be calculated by simply multiplying the average small area estimate calculated in (\ref{saeMean}) by the total area of the small area.

Despite the large improvements in estimate precision, the model-based estimates did consistently predict higher biomass relative to the design-based estimates, with the degree of positive bias varying substantially across species (Table \ref{tab:modelCor}, Figure \ref{fig:sampleSizeBias}). Bias was largest for loblolly pine, the most common and abundant species in the region (Figure \ref{fig:directCompare}c,d). Overestimation was fairly consistent within the core of loblolly pine's range, with the largest areas of overestimation occurring in regions of high loblolly pine biomass. One potential reason for this is the limited ability of the covariates in Stage 2 of the model to predict areas of high biomass for loblolly pine, which results in the spatial component of the model attempting to explain such variation and providing more smooth estimates of high biomass in the surrounding region. This phenomenon is likely the main driver of the general tendency across all species to have, on average, higher model-based estimates compared to the direct estimates. Biomass of loblolly pine and other managed species in the study area (e.g., slash pine, longleaf pine) is heavily driven by management practices that are localized in space, and thus the ability to predict fine-scale biomass variation within a model-based framework may be limited for intensively managed species, unless such information can be incorporated into the model via additional auxiliary variables. More generally across all species, inclusion of covariates that characterize forest productivity (e.g., soil characteristics, disturbance history) as well as publicly available LiDAR data may provide more accurate estimates, particularly if covariates can explain fine-scale variation not accounted for by broad climatic and environmental variables.  Additionally, use of the true FIA plot coordinates instead of the fuzzed plot coordinates used in this analysis would likely improve associations between species-specific biomass and the auxiliary variables, resulting in more accurate small area estimates. Such tasks represent important next steps to improve species-specific biomass estimates from FIA data such as those presented here.  

While the higher estimates from the proposed model relative to the design-based estimates indicate areas of improvement for the model in the FIA case study, direct estimates of species-specific county-level biomass themselves may not accurately represent true characteristics of the county. When sample sizes within counties are small, the direct estimate solely relies on the small set of plots to generate an estimate (i.e., the sample mean). Sampling theory dictates that any individual direct estimate based on a small sample size may not itself be particularly close to the true value (i.e., given the large standard error of such an estimate). In the southern US case study, 25\% of the counties contained less than 17 FIA plots, and thus such estimates could be substantially far from the true biomass density in the county. Thus, large discrepancies between model-based and design-based estimates in counties with low sample sizes does not inherently suggest inferior performance of the model-based approach in these areas, but rather represents a discrepancy between two estimates of some unknown, true population value. This concept complicates comparisons of accuracy between design-based and model-based estimates. Simulation studies such as the one implemented here provide an alternative approach for comparing model-based and direct estimators where the true underlying population is known. 

Despite the advances our proposed model provides in estimate precision, there are certain limitations that require future consideration. First, the log transformation of biomass occasionally resulted in extremely large uncertainty for the small area estimates (e.g., western Texas for live oak and water oak, Supplemental Information S1: Figures S6 and S19). Given this only occurred on the boundaries of our study region, this could be related to limitations of the isotropic and stationary exponential covariance function used to estimate the spatial processes. The use of more complex covariance functions (e.g., anisotropic functions) could help mitigate this pattern, which have been shown to provide improvements over isotropic functions in ecological applications (e.g., \citealt{finley2011comparing}). Additionally, counties along the edge of a species distribution within the study area often had more minimal precision gains relative to counties within the core of its range (e.g., laurel oak, live oak, slash pine). This could similarly reflect limitations of the isotropic exponential covariance function, or could result from localized drivers of biomass at these range edges that could not be explained by the covariates included in the analysis. Finally, care must be used when determining the size of the grid needed for generating the small area estimates \citep{finley2024models}. If the grid is not fine enough to represent spatial variation adequately throughout the small area of interest, the resulting estimates can be unstable. To ensure small area estimates are stable, we recommend predicting across multiple grid resolutions for at least a subset of the small areas of interest and comparing the resulting estimates. Because increasing the resolution of the grid will also increase the computational burden of the prediction task, one should choose the lowest resolution grid at which the estimates stabilize upon a single value. 

NFIs are increasingly being called upon to provide insights on forest resources at small spatial scales. Forest parameter estimates by species are often desired at these scales to inform wildlife habitat protection, assessments of timber supply and products, and strategic management planning. Given the sample sizes that are feasibly collected by NFIs, classic design-based estimates are not able to provide sufficient precision to meet these needs. The proposed multivariate spatial model developed here provides more precise estimates of species-specific forest parameters from NFIs, ultimately generating more opportunities for managers, landowners, and researchers to address small-scale management and research questions with NFI data. 

\section{Author Contributions}

\textbf{Jeffrey W. Doser}: Writing – review \& editing, Writing – original draft, Visualization, Validation, Software, Methodology, Investigation, Formal analysis, Data curation, Conceptualization. \textbf{Malcolm S. Itter}: Writing – review \& editing, Data curation, Methodology. \textbf{Grant M. Domke}: Writing – review \& editing, Validation. \textbf{Andrew O. Finley}: Writing – review \& editing, Methodology, Validation, Conceptualization, Supervision, Resources, Funding acquisition.

\section{Acknowledgments}

We thank two anonymous reviewers for their insightful comments that improved the manuscript. We acknowledge the computing resources provided by North Carolina State University High Performance Computing Services Core Facility (RRID:SCR\_022168). This work was supported by USDA Forest Service (FS), USDA FS/NCASI Partnership on Small Area Estimation, National Science Foundation (NSF) DEB-2213565 and DEB-1946007, and National Aeronautics and Space Administration (NASA) CMS Hayes-2023. The findings and conclusions in this publication are those of the authors and should not be construed to represent any official USDA or US Government determination or policy. 

\section{Conflict of Interest}

The authors declare no conflict of interest.

\section{Data Availability Statement}

All code and data associated with this manuscript are available on GitHub (\url{https://github.com/doserjef/Doser_et_al_2025_SAE}) and will be posted on Zenodo upon acceptance.

\clearpage

\newpage

\section*{Tables and Figures}

\begin{table}[ht!] 
  \begin{center}
  \caption{Statistics summarizing the bias and precision of model-based county-level estimates relative to the design-based direct estimates for the twenty species modeled in the FIA analysis. The Pearson correlation coefficient ($\rho$) is used as a simple measure of correspondence. Bias indicates the model-based estimate minus the design-based estimate, averaged across all counties. \% Improv. indicates the percentage of the 1,304 counties with a lower model-based CV compared to the direct estimate CV.}
  \label{tab:modelCor}
  \begin{tabular}{c c | c c c}
    \toprule
    Common Name & Scientific Name & $\rho$  & Avg & \%\\
    & &  & Bias & Improv. \\
    \midrule
    American beech & \textit{Fagus grandifolia} Ehrh. & 0.78 & 0.30 & 77.4 \\
    Black oak & \textit{Quercus velutina} Lam. & 0.79 & 0.19 & 93.6\\
    Chestnut oak & \textit{Quercus montana} Wild. & 0.93 & 0.56 & 82.7 \\
    Laurel oak & \textit{Quercus laurifolia} Michx. & 0.88 & 0.21 & 86.0 \\
    Live oak & \textit{Quercus virginiana} Mill. & 0.89 & 0.36 & 76.4 \\
    Loblolly pine & \textit{Pinus taeda} L. & 0.95 & 6.32 & 73.2\\
    Longleaf pine & \textit{Pinus palustris} Mill. & 0.72 & 0.57 & 68.9 \\
    Mockernut hickory & \textit{Carya tomentosa} Sarg. & 0.72 & 0.28 & 98.9\\
    Northern red oak & \textit{Quercus rubra} L. & 0.82 & 0.26 & 94.3\\
    Pignut hickory & \textit{Carya glabra} Mill. & 0.84 & 0.38 & 96.4 \\
    Post oak & \textit{Quercus stellata} Wang. & 0.92 & 0.49 & 91.7\\
    Red maple & \textit{Acer rubrum} L. & 0.88 & 0.18 & 97.2 \\
    Scarlet oak & \textit{Quercus coccinea} Muenchh. & 0.85 & 0.22 & 95.2 \\
    Shortleaf pine & \textit{Pinus eliottii} Engelm. & 0.91 & 0.17 & 92.6 \\
    Slash pine & \textit{Pinus echinata} Mill. & 0.96 & 0.64 & 76.4 \\
    Southern red oak & \textit{Quercus falcata var. falcata} Michx. & 0.69 & 0.39 & 94.2 \\
    Sweetgum & \textit{Liquidambar stryaciflua} L. & 0.86 & 1.13 & 97.8 \\
    Water oak & \textit{Quercus nigra} L. & 0.86 & 0.52 & 97.9 \\
    White oak & \textit{Quercus alba} L. & 0.86 & 1.97 & 96.2\\
    Yellow poplar & \textit{Liriodendron tulipifera} L. & 0.89 & 1.31 & 97.7 \\
    \bottomrule
  \end{tabular}
  \end{center}
\end{table}

\newpage

\begin{table}[ht!] 
  \begin{center}
  \caption{Comparison of model-based and design-based county-level estimates for the four-fold cross-validation assessment in the FIA analysis. Bias indicates the model-based estimate minus the design-based estimate, averaged across all counties. Values are averaged across the four-folds.}
  \label{tab:cvTable}
  \begin{tabular}{c | c c}
    \toprule
    Common Name & Correlation Coefficient ($\rho$)  & Average Bias \\
    \midrule
    American beech &  0.67 & 0.40 \\
    Black oak & 0.71 & 0.25 \\
    Chestnut oak & 0.84 & 0.53 \\
    Laurel oak & 0.80 & 0.31 \\
    Live oak & 0.77 & 0.36 \\
    Loblolly pine & 0.92 & 5.98 \\
    Longleaf pine & 0.54 & 0.68 \\
    Mockernut hickory & 0.65 & 0.35 \\
    Northern red oak & 0.79 & 0.36 \\
    Pignut hickory & 0.78 & 0.41 \\
    Post oak & 0.81 & 0.55 \\
    Red maple & 0.82 & 0.20 \\
    Scarlet oak & 0.80 & 0.24 \\
    Shortleaf pine & 0.84 & 0.33 \\
    Slash pine & 0.91 & 0.51 \\
    Southern red oak & 0.57 & 0.61 \\
    Sweetgum & 0.77 & 1.12 \\
    Water oak & 0.76 & 0.68 \\
    White oak & 0.79 & 2.03 \\
    Yellow poplar & 0.82 & 1.32 \\
    \bottomrule
    \bottomrule
  \end{tabular}
  \end{center}
\end{table}

\newpage

\newpage

\newpage

\newpage

\newpage

\end{document}